\documentclass{aastex}
\usepackage{emulateapj5}

\newcommand{\pten}[1] {\(\times\)10\(^{#1}\)}

\shorttitle{Chandra Observation of A1060}
\shortauthors{Yamasaki et al.}

\begin{document}


\title{Chandra observation of the central galaxies in A1060 cluster
of galaxies}


\author{N. Y. Yamasaki, T. Ohashi, } 
\affil{Deportment  of Physics, Tokyo Metropolitan University, 
1-1, Minami-Ohsawa, Hachioji, Tokyo 192-0397, Japan}
\email{yamasaki@phys.metro-u.ac.jp}

\and

\author{T. Furusho}
\affil{Code 662, NASA Goddard Space Flight Center, Greenbelt, MD 20771}


\begin{abstract}
{\it Chandra} observation of the central region of the A1060 cluster
of galaxies resolved X-ray emission from two giant elliptical
galaxies, NGC 3311 and NGC 3309. The emission from these galaxies
consists of two components, namely the hot interstellar medium (ISM)
and the low-mass X-ray binaries (LMXBs). We found the spatial extent
of the ISM component was much smaller than that of stars for both
galaxies, while the ratios of X-ray to optical blue-band luminosities
were rather low but within the general scatter for elliptical
galaxies. After subtracting the LMXB component, the ISM is shown to be
in pressure balance with the intracluster medium of A1060 at the outer
boundary of the ISM\@. These results imply that the hot gas supplied
from stellar mass loss is confined by the external pressure of the
intracluster medium, with the thermal conduction likely to be
suppressed.  The cD galaxy NGC 3311 does not exhibit the extended
potential structure which is commonly seen in bright elliptical
galaxies, and we discuss the possible evolution history of the very
isothermal cluster A1060\@.
\end{abstract}



\section{Introduction}
The X-ray emission from elliptical galaxies directly tells us physical
conditions of the interstellar medium (ISM), such as total mass,
thermal energy, and chemical composition. Also, the number of low-mass
X-ray binaries (LMXBs) gives us a clue how actively supernova
explosions have occurred in the past.  Systematic relation between the
X-ray and the optical blue-band luminosities, called the $L_{X}-L_{B}$
relation, was found with the {\it Einstein} observatory \citep{can87} and
has been used to study the origin of the X-ray emission.  The hard
component above $\sim 3$ keV mainly comes from the LMXBs, and its
luminosity is almost proportional to the optical one. The softer ISM
component ($\lesssim 1$ keV) shows a large scatter in the
$L_{X}-L_{B}$ relation.  \citet{mat01} analyzed PSPC data of 52
early-type galaxies and found that the galaxies are categorized into
two groups, one is an X-ray luminous group ($L_{ISM:X}>10^{41}$ erg
s$^{-1}$) characterized by extended hot halos with a radius of a few
times 10$r_{e}$, and the other is an X-ray faint group showing compact
halos.  As shown clearly
in the case of NGC 4636 \citep{mat98}, a large part of the emission in
the X-ray luminous galaxies comes from an extended ($r \sim 300$ kpc)
hot ISM, indicating that they have extended potential wells around the
galaxies with a scale as large as groups of galaxies.

In many cluster of galaxies, the central regions exhibit strong X-ray
emission from low temperature ($kT \leq 1$ keV) gases, which have been
interpreted either due to cooling flows or to ISM in central cD
galaxies.  The recent results for the cooling flow phenomena from {\it
Chandra} and {\it XMM} indicated that the simple cooling-flow picture
does not hold in a straightforward manner, and that some heating
sources must be working in the center \citep{tam01,kaa01,pet01,mol01,sas02}. 
The {\it ASCA} study of the Perseus cluster \citep{eza01,fur01} revealed 
extended ($\leq 500$ kpc) cool emission in the center,
suggesting a potential structure larger than that of the cD galaxy
NGC 1275.  The excess central emission of the Centaurus cluster shows
that the intra-cluster medium (ICM) is not isothermal but requires at least a two-phase gas
\citep{ike99}.  \citet{pao02} found that the X-ray surface brightness 
profile around NGC 1399 in the Fornax cluster had three components, i.e.\ 
cooling flow region, galactic and cluster halos.
The formation process of these hierarchical
potential structures around a galaxy or at the center of a cluster of
galaxies is not yet understood.

Interaction between the ISM and the ICM may  cause
a significant effect on the galaxy evolution.  Elliptical galaxies in
the regions of high galaxy density tend to be X-ray faint
\citep{was91}. Recently, \citet{vik01} studied the elliptical galaxies
NGC 4874 and NGC 4879 in the center of the Coma cluster and found that
the sizes of the ISM are as small as 3 kpc. 
The good spatial
resolution of the {\it Chandra} observatory enables us to look into
the interaction process around bright galaxies in a number of clusters
of galaxies.

A1060 (Hydra I cluster, $z=0.0114$) is an X-ray bright cluster of
galaxies and is considered to be the archetype relaxed system.
The {\it ASCA} and {\it ROSAT} observations of the cluster
\citep{tam96, tam00} showed that the temperature of the ICM 
is constant at $kT=3.1^{+0.3}_{-0.5}$ keV with the X-ray
luminosity $L_{\rm X}=2\times 10^{43}$ erg s$^{-1}$ in the 2--10 keV
band. An upper limit for the flux of the cool component assuming $kT=1 $
keV is $6\times10^{41}$ erg s$^{-1}$ in the energy band 0.5--3 keV\@.
The X-ray morphology of the cluster is symmetric, and the surface
brightness profile is approximated by a single $\beta$ model or a
modified ``universal'' NFW model \citep{tam00}.

There are two giant elliptical galaxies in the center of the cluster.
NGC 3311 is the cD galaxy with $m_{V}=12.65$. An E3 galaxy NGC 3309
with $m_{V}=12.60$ lies at only $1'.7$ (22 kpc in projection) away
from NGC 3311, i.e.\ within the core radius of the cluster ($3'.9\pm
0'.1$) obtained by the $\beta$ model fit for the PSPC surface
brightness \citep{tam00}.  The redshifts of the two galaxies
correspond to 3593 km s$^{-1}$ for NGC 3311 and 4075 km s$^{-1}$ for
NGC 3309, respectively \citep{RC3}.  Assuming that NGC 3311 is settled
at the center of the cluster potential as the cD galaxy, the velocity of
NGC 3309 in the ICM would be at least 500 km s$^{-1}$. The apparently smooth
and isothermal ICM and the presence of two giant galaxies makes this 
system a suitable
object for the study of interaction features between galaxies and the
ICM\@.

We use  $H_{0}=75$ km s$^{-1}$Mpc$^{-1}$ and $q_{0}=0.5$, 
where the luminosity distance to A1060 is 
46 Mpc and an angular size of $1''$ corresponds to 0.217 kpc.
The solar number abundance of Fe relative to H is 
taken to be $4.68 \times 10^{-5}$ \citep{and89} throughout the paper.

\section{Observation}
The {\it Chandra } \citep{wei00} observation was carried 
out with ACIS-I on 2001
June 4 for a total exposure time of 32 ks (OBSID: 2220). 
The observational mode
was Very Faint (VF) using 5 CCD chips (I0123, S2), and the
center of the cluster was focused on ACIS-I3\@.  The analysis was
performed on the standard processed data, which were prepared by 
the {\it Chandra}
X-ray Center (CXC), using the CIAO v2.1 package and calibration products
CALDB v2.8.  During the observation, the count rate of ACIS-I3 was
stable at $4.07\pm 0.17$ (RMS) count s$^{-1}$ when averaged in a 161 s
interval, except for a gradual increase by 20\% in the last 1.2 ks of
the whole observation.  We filtered out the data with count rates
exceeding 4.5 count s$^{-1}$, which resulted in the available exposure
time of 30463 s.

\section{Analysis}
\subsection{Image and the ICM component}
\begin{figure}
\plotone{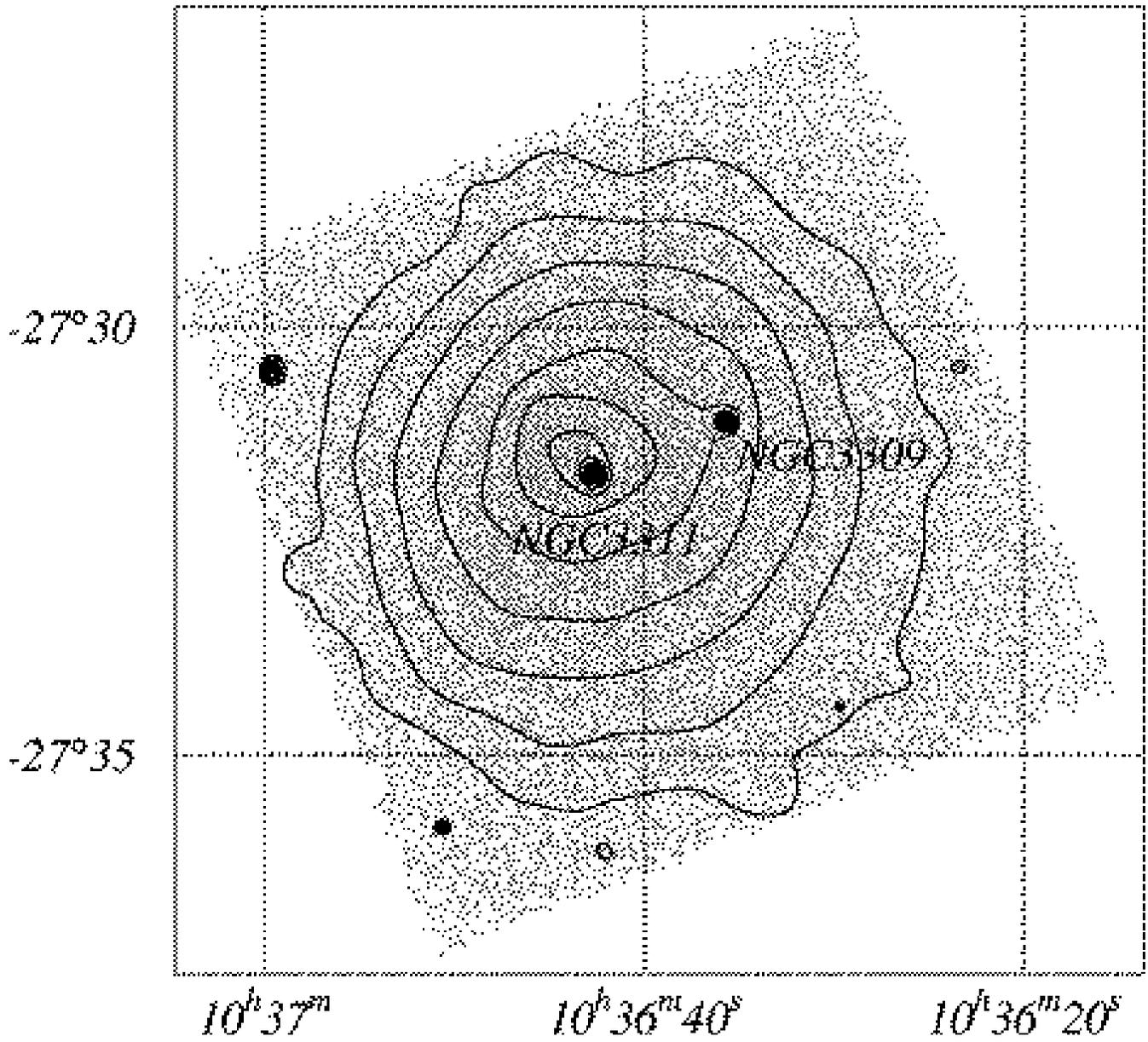}
\caption{The raw X-ray image in the energy range 0.2--1.5 keV of the
central region of A1060 taken with {\it Chandra} ACIS-I3 chip, overlaied
by adaptively smoothed contours.  The
image and contours are logarithmically scaled.  } \label{fig:image}
\end{figure}

The exposure corrected image taken with ACIS-I3 in the energy band
0.2--1.5 keV is shown in Figure~\ref{fig:image}. No background is
subtracted here.  Contours represent an adaptively smoothed image calculated 
with the CSMOOTH algorithm in the CIAO package, and 
show logarithmic X-ray intensities between 1\pten{-5} and
1.5\pten{-3} count arcsec$^{-2}$ cm$^{-2}$.
The peak near the center of the image corresponds to the cD
galaxy NGC 3311, and another peak at $1.7'$ west is NGC 3309. 
These peaks are clearly extended as compared 
with the combined response for HRMA (with 50\% encircled energy 
diameter of $0.8 ''$) \citep{jer00} and ACIS ($0.5''$), 
suggesting that the emission comes from the ISM of the
galaxies.

\begin{figure}
\plottwo{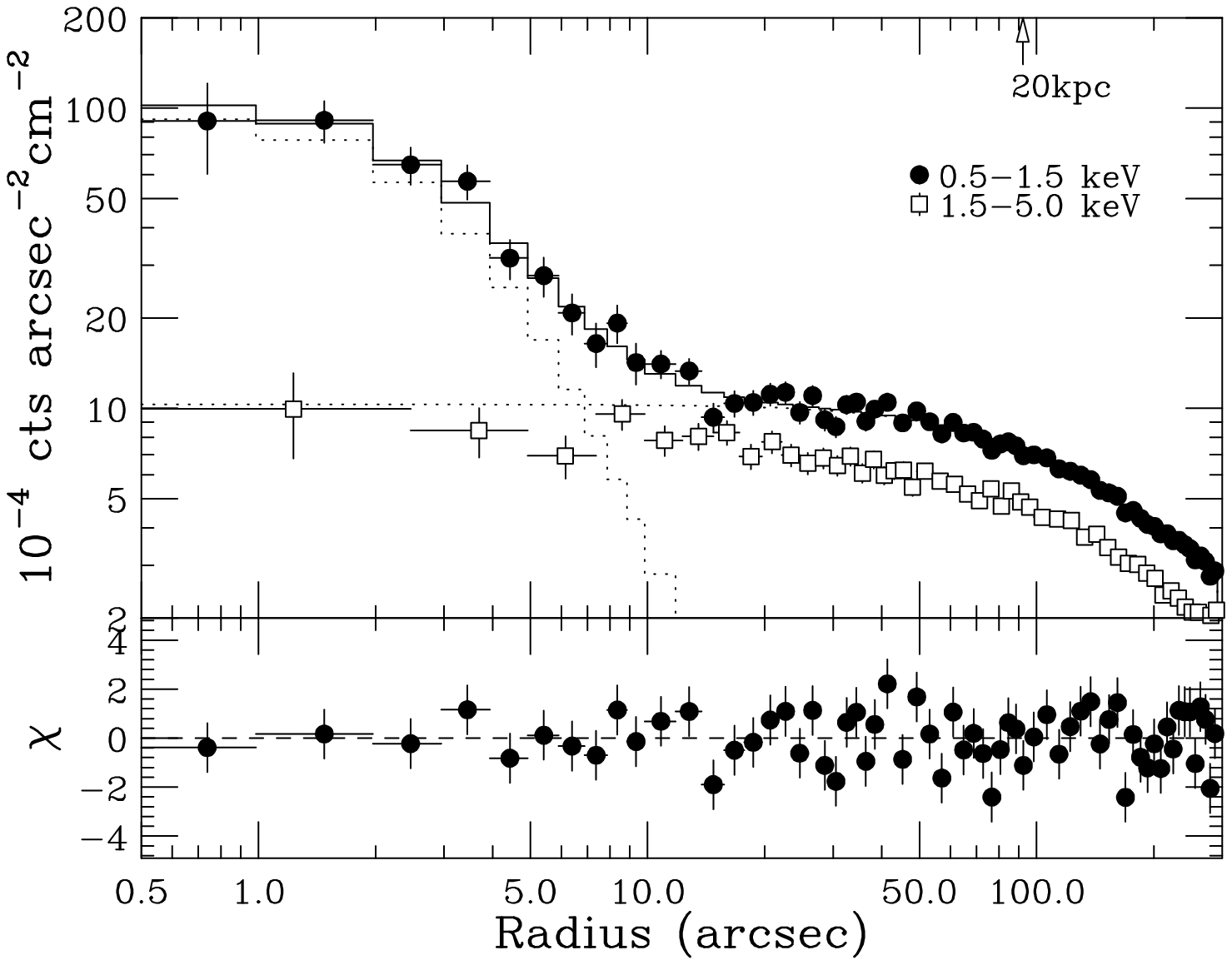}{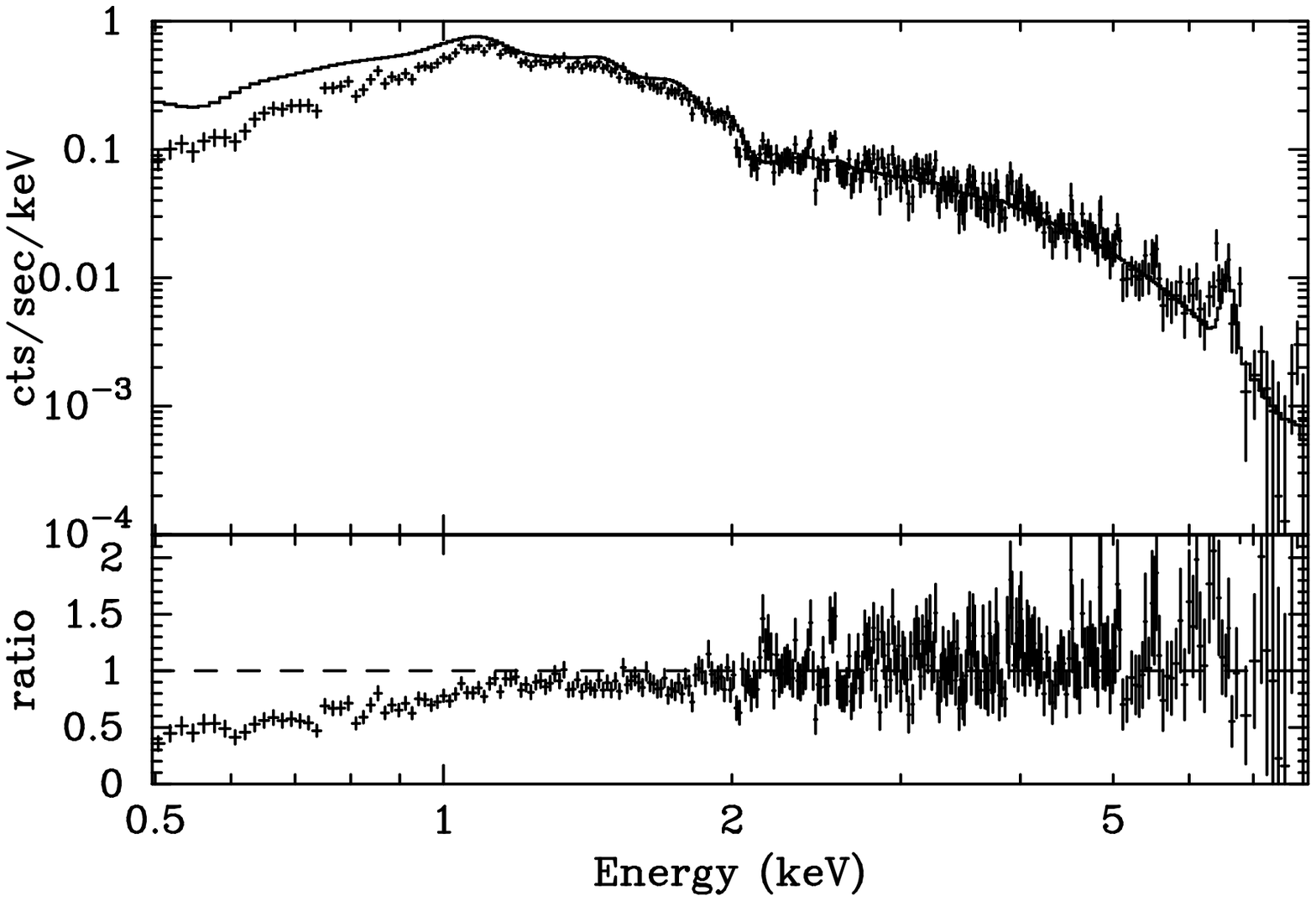}
\caption{Left: Radial profiles of the X-ray intensity 
around the center of
A1060 within a radius of $4'.9$. Filled circles are for
the energy range 0.5
-- 1.5 keV, and open  squares are for 1.5 -- 5.0 keV, respectively.
The intensity peak within $r = 10''$ in the 0.5--1.5 keV band
corresponds to NGC 3311.  The best-fit double-$\beta$ model is
overlaid on the 0.5--1.5 keV data and the bottom panel shows the 
residuals of the fit.  
Right: The pulse-height spectrum of
the ICM component ($r = 20''-98''$) compared with
the best-fit model for the {\it ASCA} data. Ratios of the data 
to the model are shown in the bottom panel.
}\label{fig:icm_radial}
\end{figure}

In the first place, we examined the radial profile of the surface
brightness.  The left panel of Figure~\ref{fig:icm_radial} shows the
profiles in the energy band 0.5 -- 1.5 keV and 1.5 -- 5.0 keV, which
include both the Cosmic X-ray background (CXB) and the non X-ray
background (NXB)\@.  The center location is asigned to NGC 3311 and the 
profile covers a
radius of $4.9'$, excluding the region with a radius of $25''$ centered on
NGC 3309. The radial profiles broadly consist of two components,
namely a narrow emission centered on NGC 3311, likely to be the ISM in
the galaxy, and an extended ICM component in A1060.  Such a
double-component structure is similar to those seen in the Fornax and
Centaurus clusters \citep{ike96,ike99,pao02}, which could be empirically
fitted with double-$\beta$ models.  We tried to fit the profile in 0.5
-- 1.5 keV with a double-$\beta$ model, which turned out to be
acceptable. The obtained parameters are listed in the first and second rows of 
Table~\ref{tab:beta}.  Looking at only the extended ICM component, the
$\beta (0.36\pm0.02)$ and the core radius ($1'.68 \pm 0'.2$) indicate
that the emission profile is very different from the previous PSPC
result characterized by $\beta=0.54 \pm 0.01$ and $r_{c}=3'.9 \pm 0.1$
\citep{tam00}.  We should note, however, that the size of the ACIS
chip ($8'.3\times 8'.3$) is comparable to the core diameter and may not
well trace the large ICM structure. This may be the cause of flatter
(small $\beta$) profile with {\it Chandra}.  
On the other hand, the PSPC data are unable 
to resolve the emission of NGC 3311, and it could well
be the reason for giving the steeper ICM structure.  We will 
deal in the cluster-scale structures of A1060 in a separate paper
(see also \citet{fur02}).

The energy spectrum of the ICM emission was examined by taking the data in 
an annulus with $r =
20''- 98'' $ excluding the NGC 3309 region, and is  shown in the right
panel of Figure~\ref{fig:icm_radial}.  The background spectrum
including the CXB and NXB is  subtracted based on the blank sky data
prepared by Maxim Markevitch (http://asc.harvard.edu/cal/). The ICM
flux is more than 10 times higher than the sum of CXB and NXB below 5
keV\@.
The spectral fit
was performed with the XSPEC v10.0 package.  For the energy response, 
we use a point source response (ARF) for the NGC 3311 as a
substitution to the one for the diffuse emission.

When the energy range is limited to 2.0--8.0 keV,
the spectral fit of the ACIS spectrum 
gives the temperature 
$kT=3.18^{+0.53}_{-0.40}$ keV
with $\chi^{2}/{\rm dof} = 262.2/215$, showing a good agreement with the
previous {\it ASCA}  and {\it ROSAT}  results, 
i.e.\ a Mekal thermal model with
$kT=3.1$ keV and metal abundance of 0.3 solar absorbed by the Galactic
column density of $N_{\rm H}=6$ \pten{20} cm$^{-2}$ \citep{tam96}.
 This suggest that the isothermal ICM is filling up a large volume in A1060, 
from the outer region to the very inner region within $r<98''$ (20 kpc).

The histogram in the right panel of Figure~\ref{fig:icm_radial} 
shows the best-fit model by \citet{tam96} 
compared with the ACIS spectrum.
Between 2 and 8 keV, the $\chi^{2}/{\rm dof}=277.6/218$ for this model. 
However, in the lower energy range down to 0.5 keV with the same 
spectral parameters, the data points
all lie under the model with the deviation by as much as a factor of 2, 
and  the $\chi^2$ value in this energy range is $2313/321$.
If we fit the spectrum by adjusting the hydrogen column density as a 
free parameter, 
the best fit value becomes 1.7\pten{21} cm$^{-2}$.
However, we are certain from the PSPC
spectrum \citep{tam00} that there is no significant absorption in
excess of the Galactic $N_{\rm H}$ (6 \pten{20} cm$^{-2}$).
The flux difference between the model and the data in 0.5--1 keV 
is 4\pten{-13} erg cm$^{-2}$ s$^{-1}$ which is about 5 times 
larger than those of the central two galaxies as shown later. 
Therefore, contamination of the galaxy component in the ASCA 
data cannot explain the spectral difference. 
We compare  the energy spectra of PSPC, GIS and ACIS from  the  same 
radius of 98'', including the galaxies NGC 3311 and NGC 3311, 
and the  descrepancy at the low energy band still exists. 
The degradation in the detection efficiency of the ACIS
below 1 keV has been recently reported 
(http://cxc.harvard.edu/cal/Links/Acis/acis/Cal\_projects/index.html), 
then this deficiency of the soft X-ray flux in the {\it Chandra} data 
 could be caused by this phenomena.

\subsection{Energy Spectra of NGC 3311 and NGC 3309}
\begin{figure}
\plottwo{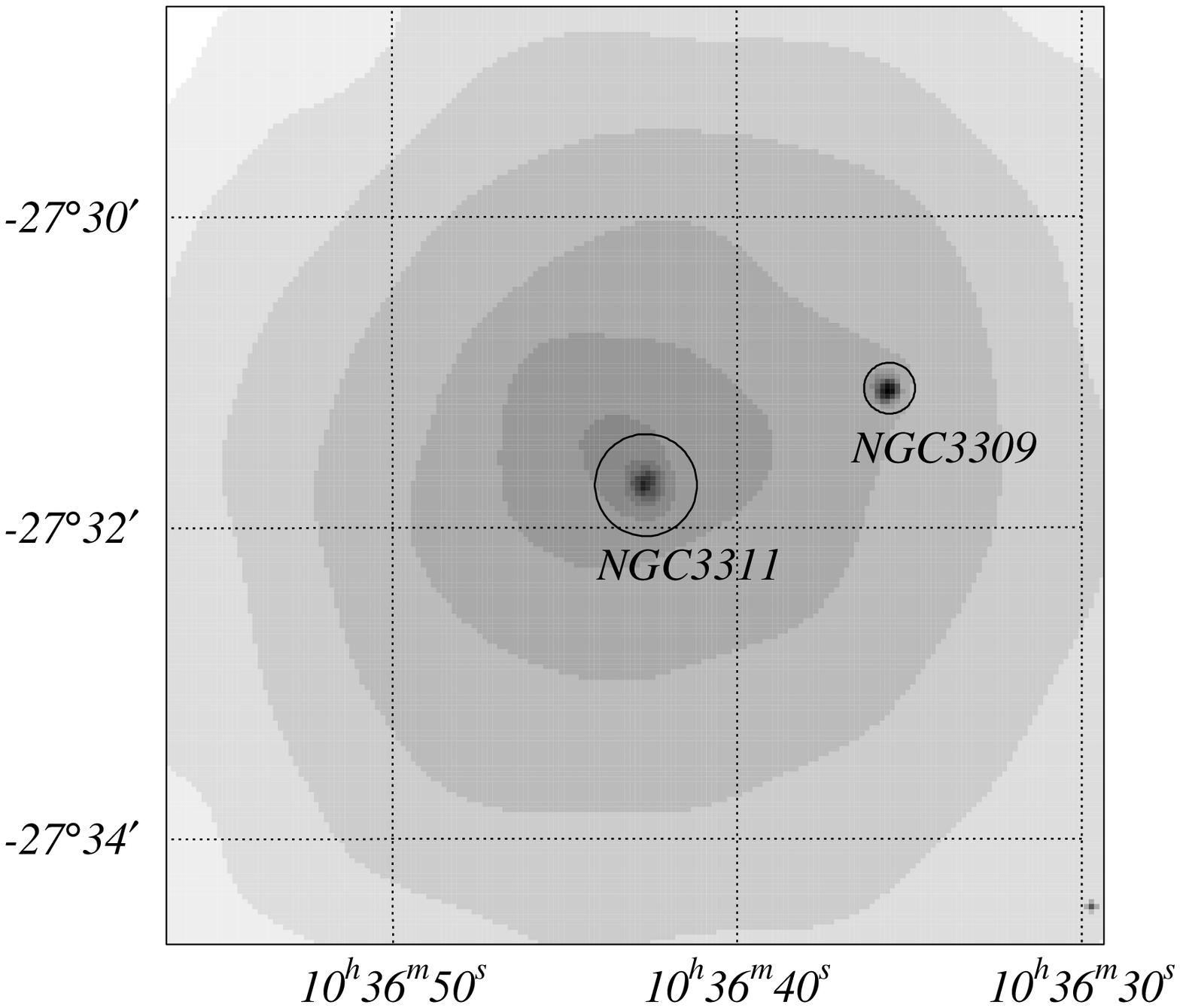}{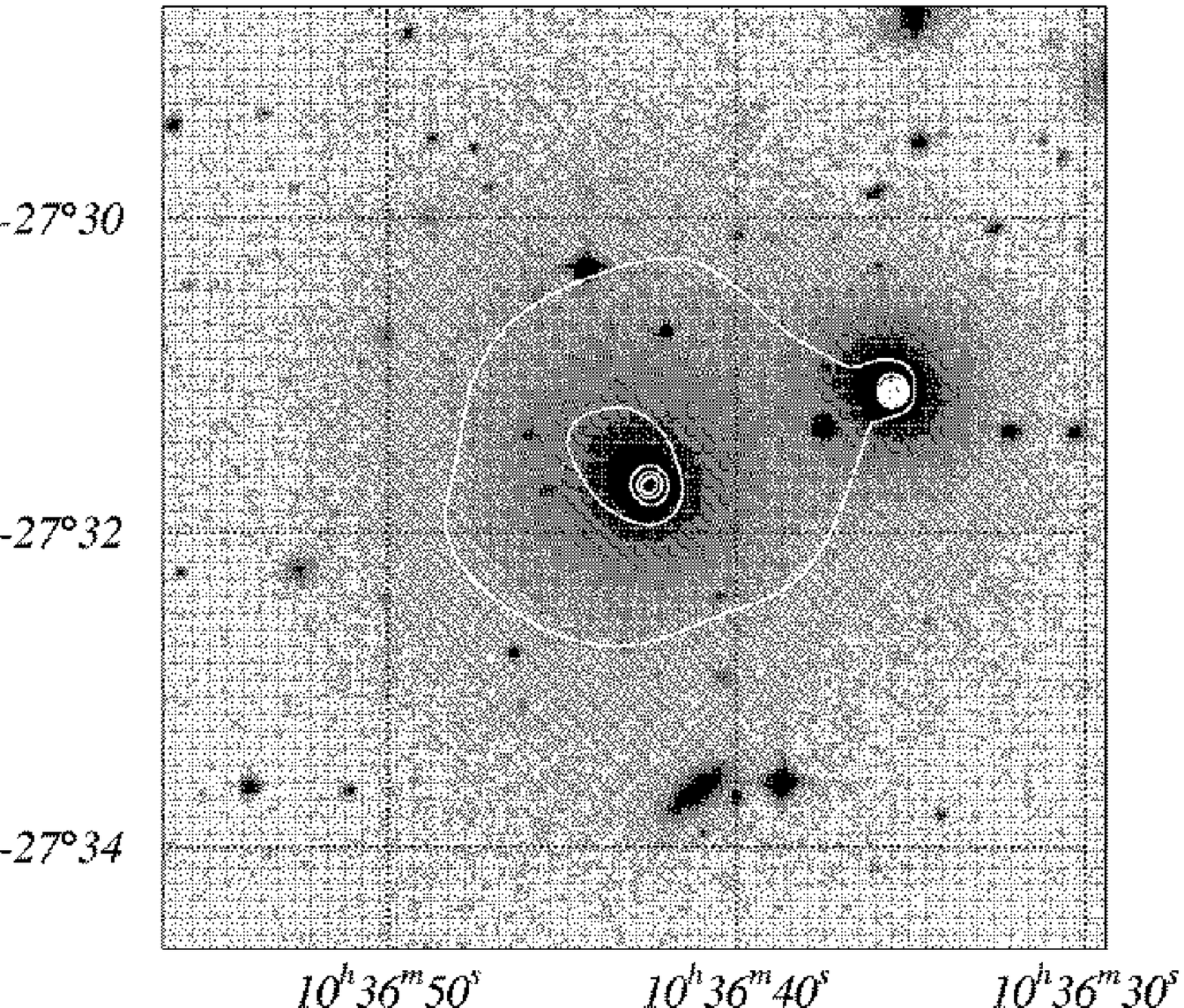}
\caption{Left: The adaptively smoothed {\it Chandra} X-ray image 
of the central $6'\times 6'$ region of A1060\@ .  
Circles indicate the integrated regions in which energy spectra
are accumulated.
Right: the corresponding optical image from the
digitized sky survey overlaid with X-ray intensity contours.  
Contours are spaced by a factor 1.4, with the lowest one at 
8.0\pten{-4} count arcsec$^{-2}$ cm$^{-2}$.
Both images are logarithmically  scaled.
}\label{fig:X_opt_image}
\end{figure}

The X-ray image including the two  central galaxies 
was compared with an
optical image from Digitized Sky Survey (DSS) for the central $6'
\times 6'$ region, as shown in Figure~\ref{fig:X_opt_image}.  
The X-ray extents of the two galaxies NGC 3311 and NGC 3309 are much
smaller than those in the optical band. This feature is quite
similar to the recent results for the two galaxies in the Coma cluster 
\citep{vik01}. The
detailed radial profiles will be analyzed in the next section.

\begin{figure}
\plottwo{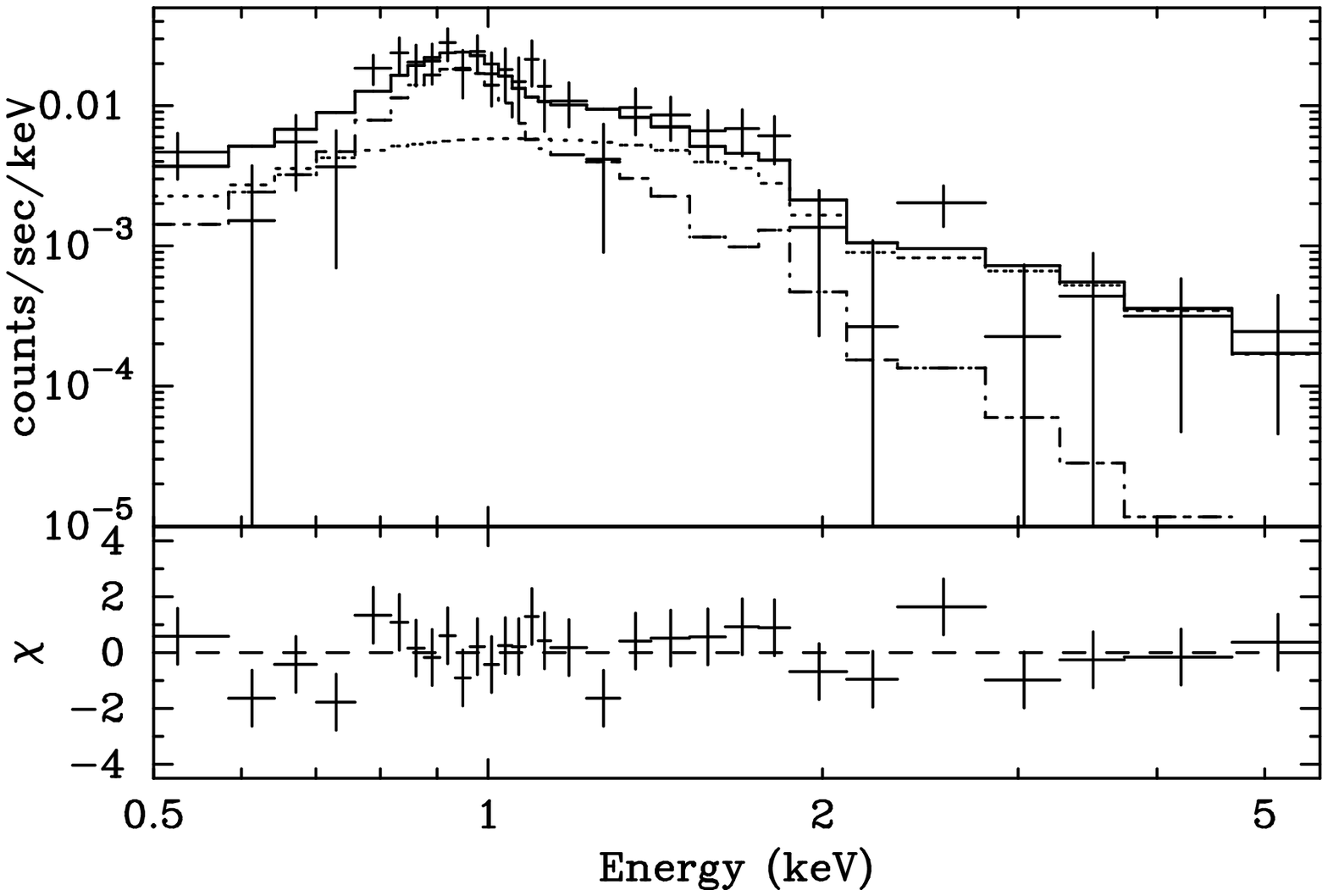}{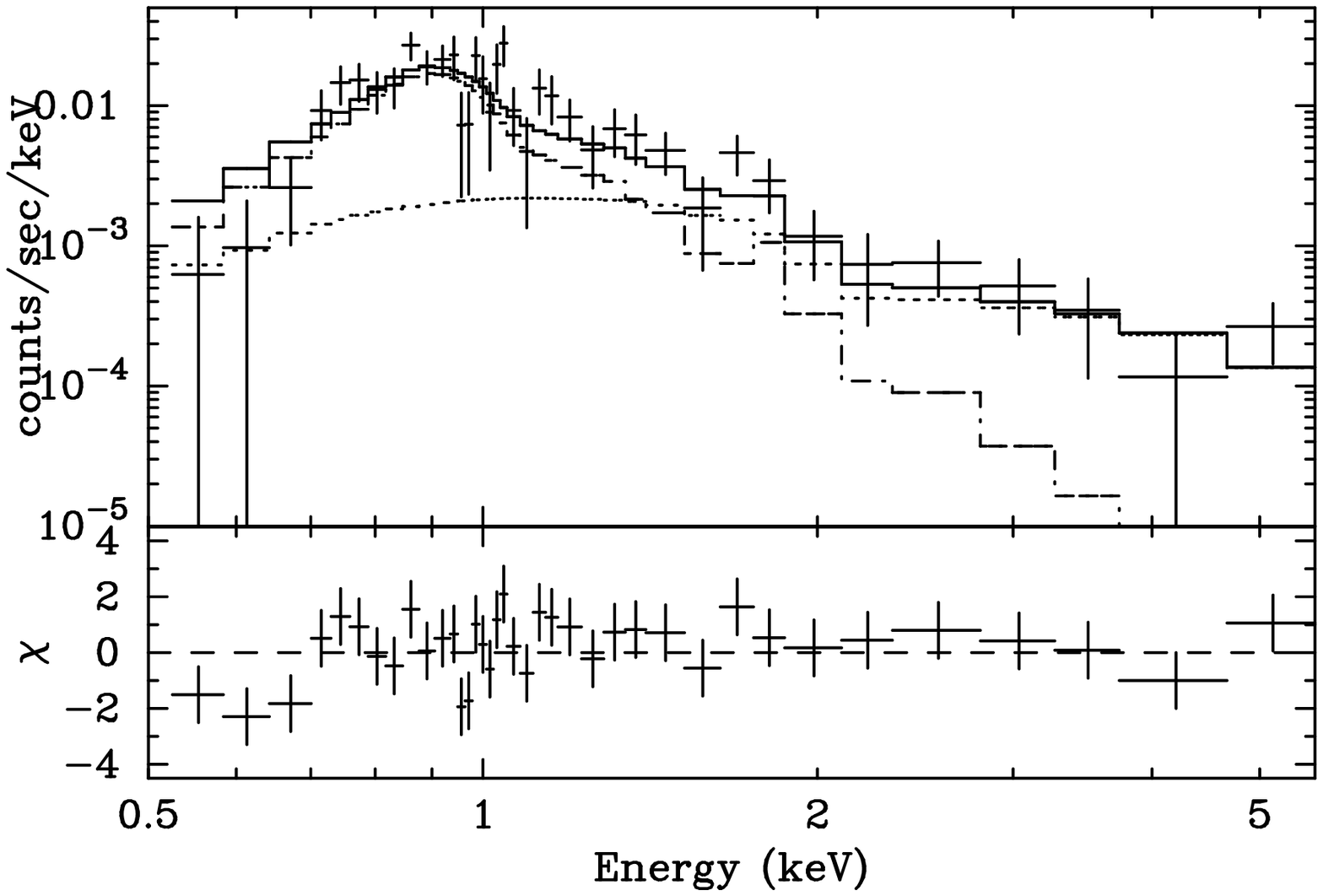}
\caption{The observed pulse-height spectra of NGC 3311 (left) and NGC
3309 (right), with the background subtracted, along with the best-fit
spectral models. The dot-dashed lines show the estimated ISM
component described by a Mekal thermal model (see Table 2), and the
dotted lines show LMXB component approximated by zbrems model with
$kT=$ 10 keV\@.  The bottom panels show residuals of the spectral fit.
}\label{fig:spectra}
\end{figure}

In this section, energy spectra of the central two galaxies, NGC 3311
and NGC 3309, are examined. The X-ray data in circular regions with
radii $20''$ and $10''$ centered on the respective galaxies were cut
out, as indicated with solid circles in
Figure~\ref{fig:X_opt_image}. These radii correspond to $0.21 r_{e}$
for NGC 3311 and $0.46 r_{e}$ for NGC 3309, respectively
\citep{vas91}.  For the common background, we took data from a
circular region with $98''$ radius in the cluster center, excluding
the two galaxies.  The resultant background spectrum is the same as that
already shown in Figure~\ref{fig:icm_radial}.  Normalization of this
background spectrum, which includes NXB and CXB, was then scaled to match the
surface brightness of the ICM at the target positions based on the radial
profile fits in the next section. Since the background spectrum is
dominated by the ICM emission, the wrong scalings for the NXB and CXB 
intensitis cause an influence which is less than 1\% of the total background
flux. The background subtracted spectra of the two galaxies are shown in
Figure~\ref{fig:spectra}.

Spectral fits were also 
carried out for the data of two galaxies with XSPEC v.10.0.
  As for the
model spectra, the Mekal model was used for the ISM component and a
zbrems model, representing a redshifted thermal bremsstrahlung
emission, was applied for the LMXB component, respectively.  The
temperature of the zbrems model was fixed to $kT = 10 $ keV, as an
empirical approximation of the X-ray spectra of LMXBs
\citep{mak89,mat94}.   The absorption and metal abundance in the Mekal
model were not effectively constrained. We, therefore, fixed the
absorption column to the Galactic value of 6 \pten{20} cm$^{-2}$ and
the metal abundance to 0.5 solar, respectively.  The resultant
parameters of the spectral fits are listed in Table~\ref{tab:spec}.
The ISM temperatures are $0.87^{+0.21}_{-0.09}$ keV and
$0.77^{+0.10}_{-0.07}$ keV for NGC 3311 and NGC 3309, respectively.

The sum of the ISM fluxes for the two galaxies is ($7\pm 2$) \pten{-14}
erg cm$^{-2}$s$^{-1}$, which is consistent with the previous upper
limit for the cool component ($kT\sim 1$ keV), 4 \pten{-13} erg
cm$^{-2}$s$^{-1}$, obtained by {\it ASCA} and PSPC
\citep{tam96,tam00}.  When the metal abundance was varied between
$Z=0.2$ and 1.0 solar, the best-fit temperature varied only by 0.02
keV but the emission measure showed a change by a factor of about 2. This is
because the emission measure goes almost inversely proportional to the
metal abundance.  The total luminosities for the sum of ISM and LMXBs
in the energy range between 0.5 and 4.5 keV are 2.4 \pten{40} erg
s$^{-1}$ and 1.6 \pten{40} erg s$^{-1}$ for NGC 3311 and NGC 3309,
respectively. The LMXB emission accounts for 50\% and 53\% of the
X-ray luminosities for the respective galaxies.

\subsection{Radial profiles of NGC 3311 and NGC 3309}

\begin{figure}
\plottwo{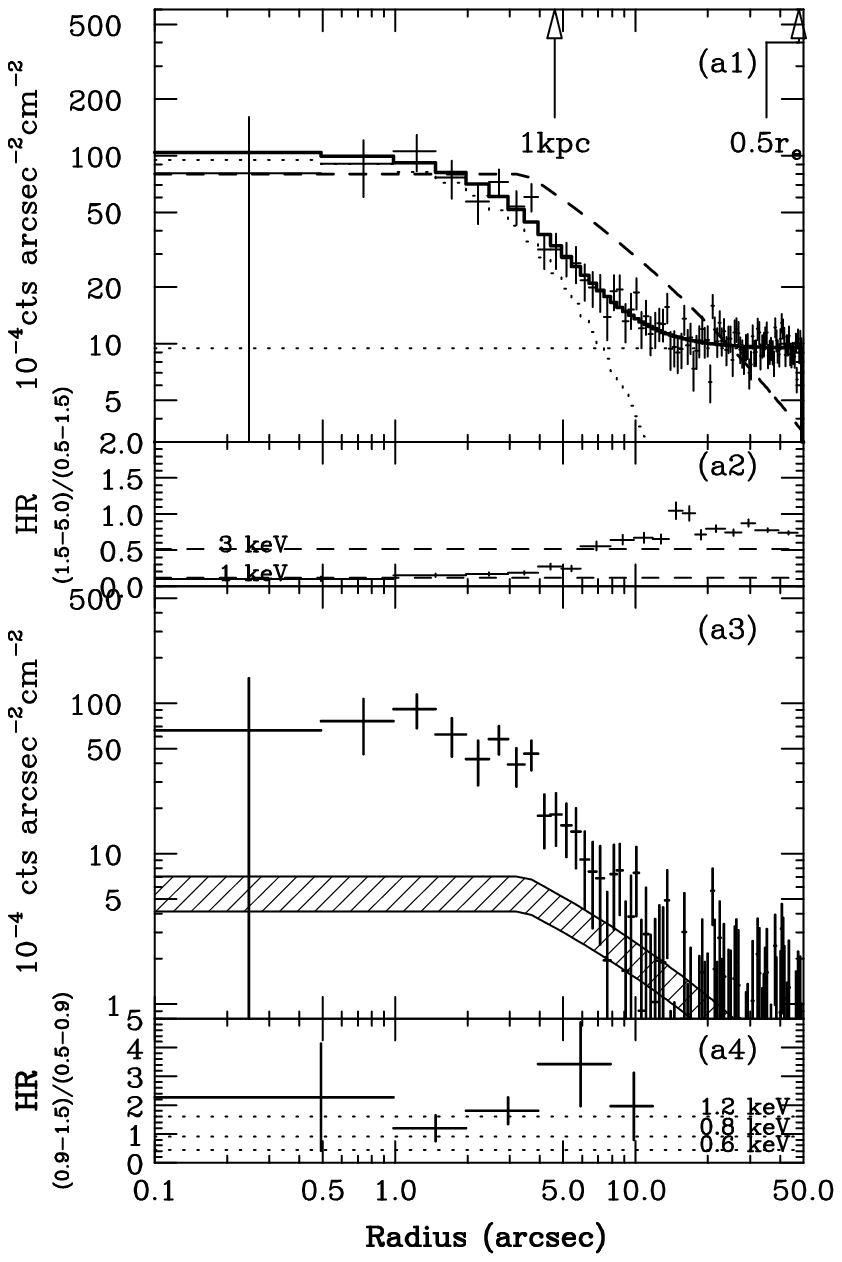}{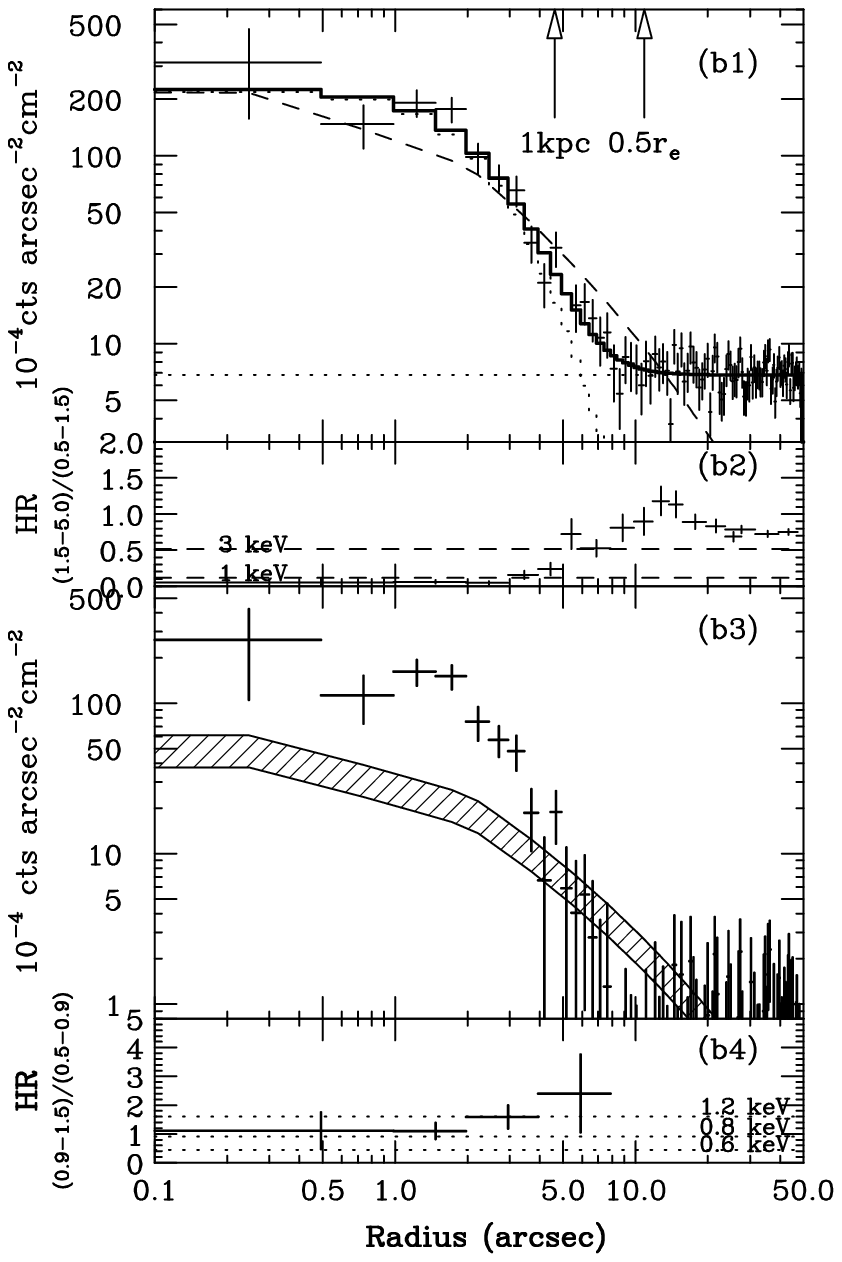}
\caption{The radial profiles of various quantities for NGC 3311 (left)
and NGC 3309 (right) within a radius of $49''$.  The top row, (a1) and
(b1), shows radial intensity profiles between 0.5 and 1.5 keV before
the background subtraction fitted with the best-fit $\beta$ model
and a constant component (dotted curves). The dashed lines 
show the optical profiles.  The second row,
(a2) and (b2), shows the hardness ratios with
the corresponding temperatures indicated with dashed
lines. In the third row, (a3) and (b3), crosses show the background
subtracted ISM intensity in the energy band 0.5--1.5 keV after
subtraction of the ICM and LMXB components. The estimated LMXB
contributions are also shown with hatched regions.  The bottom row, (a4)
and (b4), shows hardness ratios for the pure ISM component, with the
corresponding temperatures plotted with dotted lines.
}\label{fig:radial}
\end{figure}

Radial profiles of the X-ray emission from the central two galaxies,
NGC 3311 and NGC 3309, are examined in some detail. The profiles
before background subtraction in the energy range 0.5 -- 1.5 keV are
shown in Figure \ref{fig:radial} (a1) and (b1) within a radius of
$49''$ for NGC 3311 and NGC 3309, respectively.  \citet{vas91}
obtained the optical surface brightness profiles with a resolution of
$1''.5$ (FWHM)\@. These optical profiles are well represented by de
Vaucouleurs law for $ r = 1 -4$ kpc (NGC 3311) and for $0.6 - 5$ kpc
(NGC 3309), respectively. The isophotal radii $r_{e}$ are $95''.5$and
$21''.8$, respectively.  The dashed lines indicate the optical surface
brightness profiles \citep{vas91}. The X-ray emission profiles are
clearly sharper than those in the optical band.  The X-ray profiles
can be fitted with single $\beta$ models and a constant
background. For the radial profile fittings, we use CERN library
package (http://wwwinfo.cern.ch).  Thick solid lines in
Figures~\ref{fig:radial} (a1) and (b1) indicate the best-fit models,
which give $\chi^{2}/{\rm d.o.f.}= 96.9/96$ and $95.6/96$ for NGC 3311
and NGC 3309, respectively. The parameters of the profile fits are
listed in the third and forth rows in Table~\ref{tab:beta}, and the
values for NGC 3311 are consistent with the previous joint fit
including the ICM in section 3.1\@.  Both galaxies indicate core radii
of about $4''$ or 1.3 kpc, which is a typical value for elliptical
galaxies \citep{for85}.  The $\beta$ value ($1.08^{+0.85}_{-0.27}$)
for NGC 3309 is, however, much larger than the average value of 0.5
obtained for nearby bright elliptical galaxies with {\it Einstein}
\citep{for85}, indicating that the emission of A1060 galaxies falls 
very sharply.  We
also tried to fit the profile for NGC 3309 with a cut-off $\beta$
model by fixing $\beta=0.5$, but the fit turned out to be unacceptable
with $\chi^{2}/{\rm d.o.f.}= 110.2/96$.  These profile fits confirm
that the X-ray emission from the two galaxies are compact and
contained within a small radius of $10'' - 20''$.

Panels in the second row in Figures~\ref{fig:radial} show hardness
ratios (HRs), which are defined as intensity ratios between energy
bands 1.5 -- 5 keV and 0.5 -- 1.5 keV\@. Since the background is not
subtracted, the fluxes contain contribution from ISM, LMXB, ICM and
NXB\@. The HR profile indicates that in the inner region ($r < 5''$),
the temperature is about 1 keV corresponding to the ISM emission and
gradually increase to the outer region.  The HR profiles show that the
temperature reaches a constant value in the outer region at $ r > 20''$. 
The HR of the ICM as shown in the energy spectrum in Figure \ref{fig:icm_radial}
is $0.77\pm 0.01$, which corresponds to about $kT \sim 3$ keV. 
We also note that the HR
profiles of both galaxies show peaks at $ r=10''-20''$ 
which are significant at the 95\% confidence limit (more than $2\sigma$ level).
This feature
is most naturally explained in terms of the emission from LMXBs, which
produce harder spectrum ($kT \sim 10$ keV) than both ISM and
ICM \citep{mat94}. Therefore, the HR profiles indicate that the 
dominant source of
the X-ray emission shifts from ISM to LMXBs at $r \sim 10''$, and then
into ICM at $r \sim 20''$.

Based on these results, we examined the radial profiles of the ISM
component only.  The surface brightness of the NXB and the ICM could
be regarded as a constant in the galaxy scale as shown in the previous
$\beta$ model fit.  As for the LMXB emission, we assume that the
spatial distribution follows the optical surface brightness profile
\citep{vas91} and their energy spectra is given by a bremsstrahlung emission 
of $kT = 10$keV\@. 
This enables us to estimate the LMXB luminosity as a function
of radius as shown by hatched curves in Figure~\ref{fig:radial} (a3)
and (b3). We included the errors due to the flux of the LMXB emission, 
the temperature and the abundance of the ISM emission.
After subtracting the contributions from LMXB, ICM and NXB in
this manner, we can obtain the ISM profile as the remaining
feature. We present the radial ISM profiles in the energy band 0.5 --
1.5 keV for NGC 3309 and NGC 3311 in Figure~\ref{fig:radial} (a3) and
(b3), respectively.

The radial profiles in Figure~\ref{fig:radial} (a3) and (b3) clearly
indicate that the sizes of the ISM emission are as small as $2 \sim 3$
kpc. If we fit the ISM profiles with a $\beta$ model, the obtained
$\beta$ values are $0.94^{+1.47}_{-0.34}$ for NGC3311 and
$1.91^{+0.92}_{-1.03}$ for NGC3309, respectively.  The large $\beta$
value for NGC 3309 suggests that the ISM is not in a simple condition
characterized by an isothermal gas under a hydrostatic equilibrium in
a King shape gravitational potential.  Based on these $\beta$ model
fits, the central gas densities are estimated as $9.4^{+6.9}_{-3.4}$
\pten{-2} cm$^{-3}$ for NGC 3311 and $1.54^{+1.12}_{-0.42}$
\pten{-1}cm$^{-3}$ for NGC 3309, respectively.  Hardness ratios for
the ISM components, defined by the ratio of the counts 
between  0.9 --1.5 keV and 0.5 -- 0.9 keV bands, are also shown as a function of radius
in (a4) and (b4).  Within the error, the ISM temperatures are
consistent to be constant with radius up to about $10''$ for both NGC
3311 and NGC 3309.
As we show later in the discussion, the LMXB luminosities 
in NGC 3311 and NGC 3309 are relatively higher than 
an X-ray faint galaxy NGC 4697 \citep{sar01}. 
But even if  we reduce the LMXB luminosity to the
half of the best-fit value , the change in the hardness ratios 
are less than 10 \% or 0.05 keV.
However, at the same time we cannot reject a model that
the ISM temperature gradually goes up with radius from the central
$\sim 1$ keV and reaches about 3 keV ($\approx$ ICM value) at $r \sim
10''$, because of the large error.

\section{Discussion}
The present {\it Chandra} observation of A1060 showed that the ICM
temperature within $98''$ from the center was consistent with the
previous {\it ASCA} and {\it ROSAT} results, $kT\sim 3.1$ keV\@. The
spectrum showed no temperature drop or additional cool component
except for the ISM emission from the two elliptical galaxies. The surface
brightness profile clearly indicated two components, i.e.\ the
spatially narrow emission confined within the cD galaxy NGC3311 and
the widespread ICM component, with the latter well approximated by a
simple $\beta$ model with $\beta=0.36$ and core radius $=1.7'$.  
Properties of the ICM distribution over the
large scale will be studied in our next paper (see also \citet{fur02}).

The X-ray emission from the ISM of two giant elliptical galaxies,
NGC3309 and NGC3311, was clearly resolved for the first time.  Their
spatial extents are as small as $10'' - 20''$, or less than $0.5r_{e}$.  
According to the recent X-ray classification of elliptical galaxies
by \citet{mat01}, these sizes are very small even among the X-ray
faint and compact class. This feature is very similar to the recent
{\it Chandra} measurement of NGC 4874 and NGC 4889 in the Coma cluster
\citep{vik01}. The X-ray sizes of these Coma galaxies are also
unusually small, and the authors discuss that the ISM pressure is
balanced with the ICM level with little sign of thermal conduction at
the outer boundary of ISM\@. 

Before looking into the origin of this compactness in the X-ray
emission region, we first evaluate the luminosity and distribution of
the LMXB component.  \citet{sar01} studied X-ray to optical
luminosity ratios of LMXBs in elliptical galaxies based on the data
from {\it Chandra} and {\it XMM}\@.  For an X-ray faint elliptical 
galaxy NGC 4697, they obtained $L_{X}/L_{B}=
8.1\times 10^{29}$ erg s$^{-1} L_{B\odot}^{-1}$ by taking the LMXB
luminosity in the energy range 0.3-10 keV\@.  
For NGC 3311 and NGC 3309, the same ratios for the entire emission
region are higher at 2.9 \pten{30} and 1.3
\pten{30}erg s$^{-1} L_{B\odot}^{-1}$, respectively.  
The reason why A1060 galaxies indicate higher $L_{X}/L_{B}$
for LMXBs can be considered in the following
way.  First, since \citet{sar01} summed up the fluxes of only
detected sources, the faint LMXBs with fluxes less than the detection 
limit were not
included.  Second, there may be some influence from nuclear radio sources
for NGC 3311 and NGC 3309.  \citet{lin85} reported a weak radio source 
in NGC 3311 and a radio jet from
NGC 3309, whose equipartition energy was 5 \pten{53} erg s$^{-1}$.
In the X-ray data, we cannot exclude possible presence of
low-luminosity AGNs with $L_{X} \leq 10^{39}$ erg s$^{-1}$
in the center of these galaxies. 
Thirdly, the possible star-formation activity in
NGC 3311 would yield a high LMXB density.  The emission line spectrum
from NGC 3311 suggests presence of H II regions \citep{vas91}, and the
metal abundance in globular clusters in NGC 3311 is extremely high
\citep{sec95} implying a young star population.  Recent HST
observation, however, obtained normal globular cluster colors and
metallicities \citep{brodie00}, which has made the high metallicity of
NGC 3311 rather uncertain.

\begin{figure}
\plotone{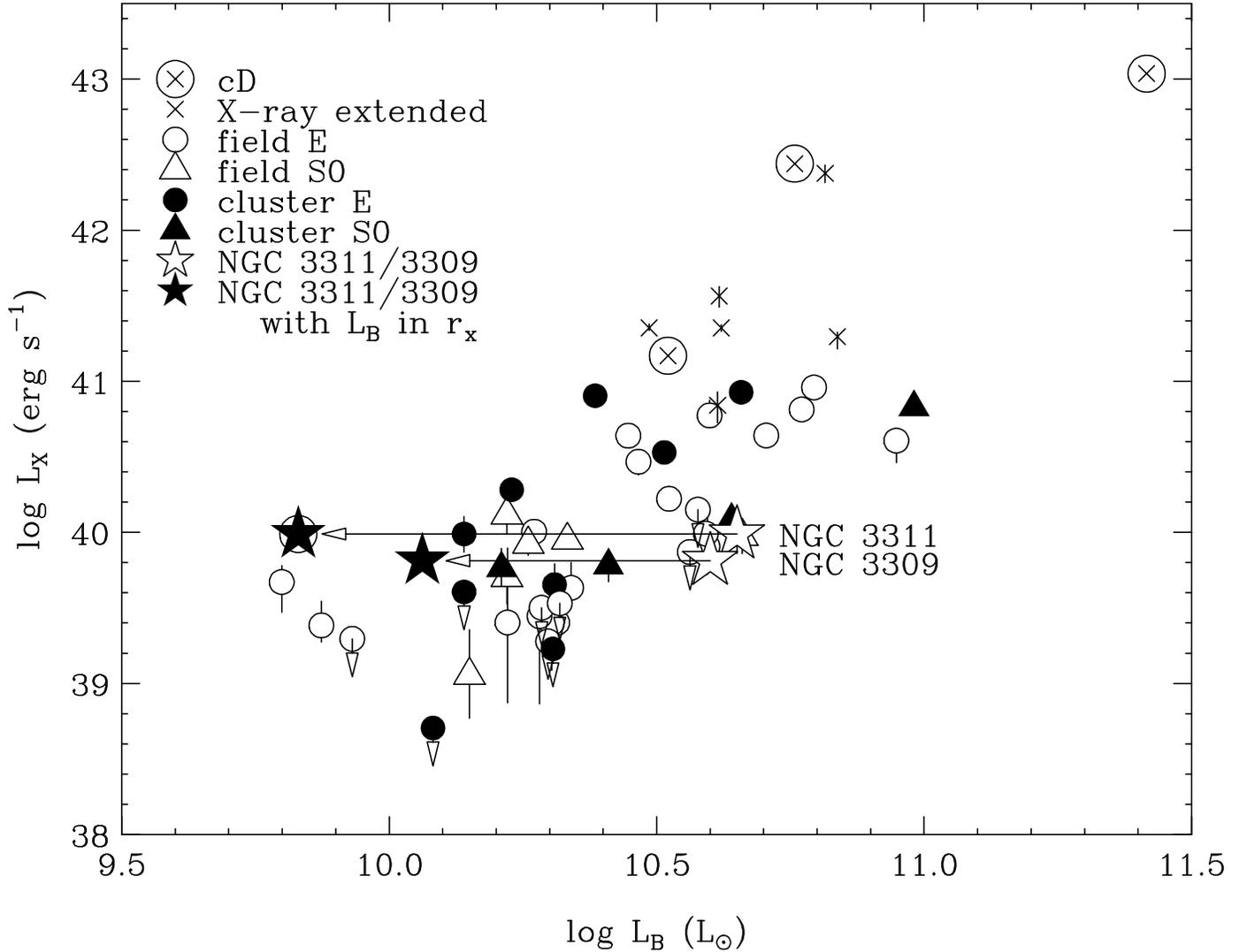}
\label{fig:lxlb}
\caption{The $L_{X}-L_{B}$ relation for elliptical galaxies based on
figure 4 of \citet{mat01}, with an addition of NGC 3311 and NGC 3309
points shown with pentagrams. $L_{X}$ values for the 2 galaxies are
calculated within the X-ray integrated radii ($0.21r_{e}$ for NGC 3311
and $0.46r_{e}$ for NGC 3309, respectively). Open pentrams are for the
total $L_{B}$, and filled pentagrams are for the $L_{B}$ only within
the X-ray integrated radii.  $L_{X}$ for other galaxies are X-ray
luminosities of the ISM in the energy range between 0.5 and 4.5 keV
within $4r_{e}$, however the values of NGC 3311 and NGC 3309 do not
change by extending the integration radii since their X-ray sizes
are compact. Symbols indicate galaxy categories, which are X-ray
extended galaxies (crosses), cD galaxies (large open circles), field
ellipticals (open circles), field S0s (open triangles), cluster
ellipticals (filled circles), and cluster S0s (filled triangles),
after \citet{mat01}.}
\end{figure}

\begin{figure}
\plottwo{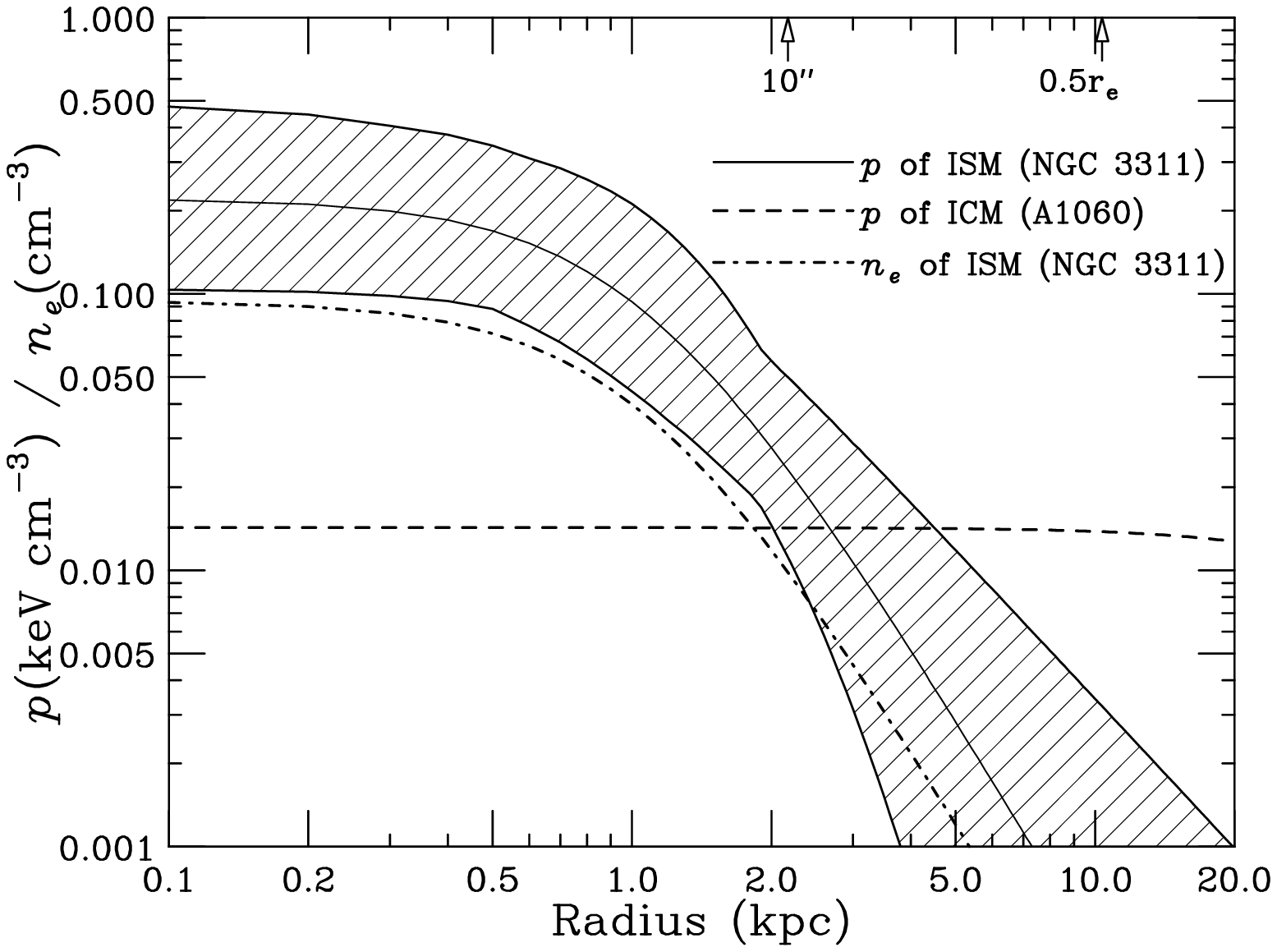}{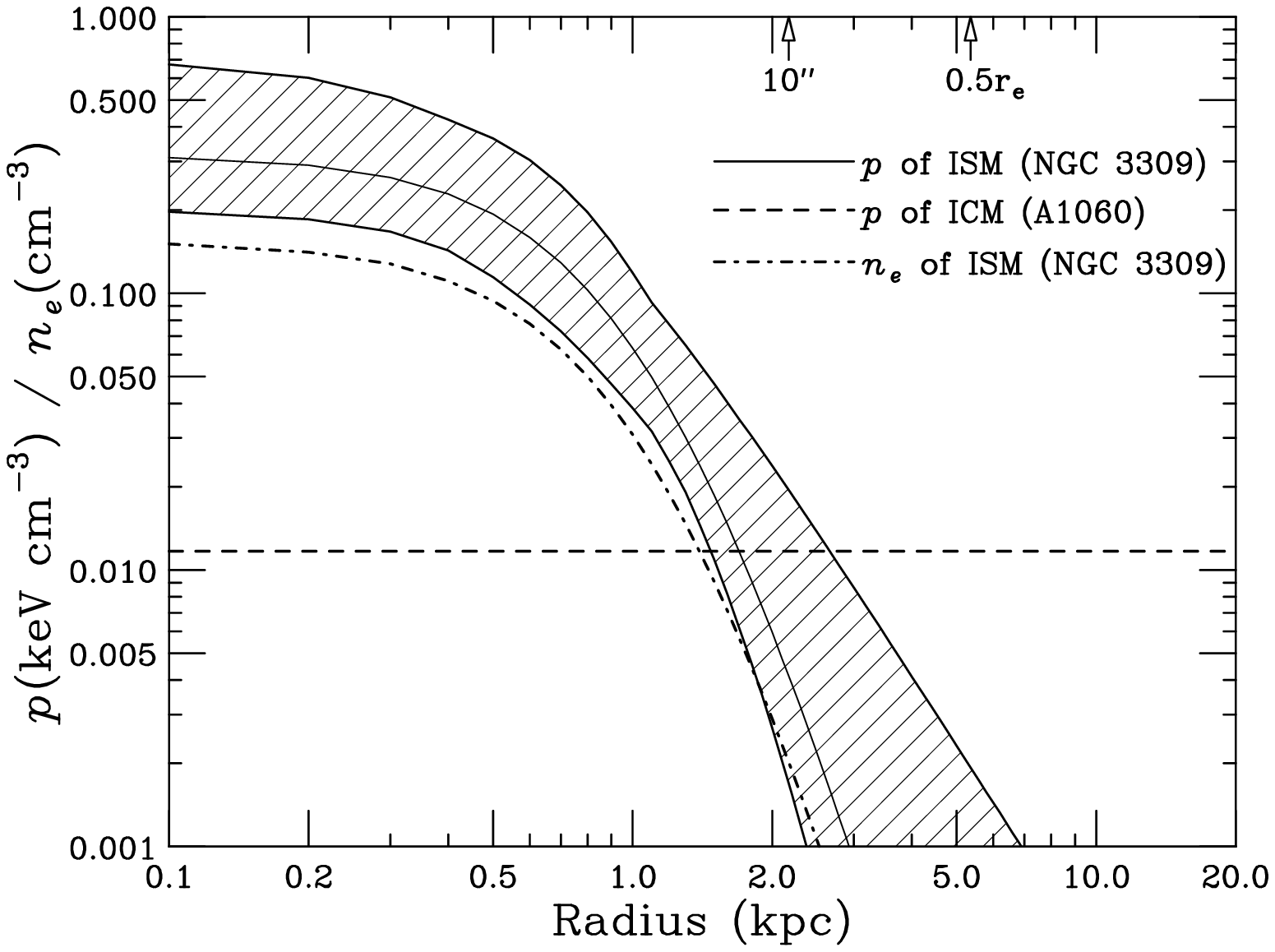}
\caption{Radial profiles of the pressure and the electron density 
for the ISM in NGC 3311 (left) and
NGC 3309 (right). 
Solid curves in the middle of the hatched regions
indicate the best-fit pressure profiles 
and the hatched regions show the 90\% error range
due to uncertainties in the $\beta$ model fit and in the emission
measure. The dot-dashed lines indicate the best-fit density profiles.
The surrounding ICM pressure is shown with the dotted line, where
we assume the density profile of the A1060 ICM given by
\citet{tam96}.
}\label{fig:pressure}
\end{figure}

\citet{mat01} derived $L_{X}/L_{B}$ relation for the ISM of 27
galaxies. In Figure \ref{fig:lxlb}, we reproduced Figure 4 of
\citet{mat01} and plotted the $L_{X}-L_{B}$ 
data of NGC 3311 and NGC 3309.  If we take the optical luminosity 
for the whole galaxy,
the two galaxies, shown with open pentagrams, indicate
slightly lower $L_{X}$ values than other X-ray compact
early-type galaxies. The X-ray integration radii used in the spectral
fits correspond to $0.21r_{e}$ for NGC 3311 and $0.46r_{e}$ for NGC
3309, respectively.  Fractions of the optical luminosity encircled by
these radii are 15.0\% and 28.6\% for the respective galaxies. If
optical luminosities only in these regions are taken,
$L_{X}/L_B$ ratios become higher by a factor of 4--6 as shown with
filled pentagrams in Figure \ref{fig:lxlb}.  These ratios are higher
than most of other galaxies. The spatial extents of the hot ISM in NGC
3311 and NGC 3309 are certainly smaller than those of stars,
nevertheless the X-ray luminosities satisfy the average $L_X/L_B$
relation.  
We note that the X-ray luminosities derived in this paper 
could be underestimated by a factor of about two at maximum 
due to the uncertainity of the quantum efficincy of the ACIS 
at the low energy band. However, the tendency described here 
is not affected.
A possible explanation would be that, even though
a substantial fraction of the ISM is stripped off,
the remaining gas is confined in the galaxy by the
external pressure of the ICM\@.  In Figure~\ref{fig:pressure}, we
plotted radial profiles of the ISM density $n_{e}$ (cm$^{-3}$) and
the pressure $\frac{3}{2}(n_{e}+n_{i})kT \sim 2.69n_{e}kT$ (keV cm$^{-3}$),
compared with the ICM pressure \citep{tam96}.  Clearly, the
pressure balance between the ISM and the ICM is reached just around
the outer boundaries of X-ray emission: 2.8 kpc and 1.7 kpc for NGC
3311 and NGC 3309, respectively.

As for the supply of the ISM matter, the total mass of the ISM is
estimated to be 5.6 \pten{7} $M_{\odot}$ for NGC 3311 and 1.1 \pten{7}
$M_{\odot}$ for NGC 3309, respectively. Under a typical stellar mass
loss rate of
$\dot{M}_{\star}=1.5\times10^{-11}M_{\odot}$yr$^{-1}L_{\odot}^{-1}$
\citep{fab76}, the ISM can be accumulated in 2.7 \pten{8} yr and 9.3
\pten{7} yr for the two galaxies. The radiative cooling times,
$t_{\rm cool}=3M_{\rm ISM}kT/\mu m_{p}L_{X}$, are 5.9 \pten{8} yr for 
NGC 3311
and 1.4 \pten{8} yr for NGC 3309.  Therefore, the stellar mass loss,
together with the input from supernova, can compensate the 
radiative energy loss. However, if thermal
evaporation works at the rate of the classical estimation by
\citet{cow77}, the ISM would be completely lost in only a few times
$10^{6}$ yrs. The presence of the X-ray halo shows this  is not
the case.  In spite of the large radial velocity of NGC 3309, higher than
the cluster average by $\sim 500$ km s$^{-1}$, its X-ray image is
circularly symmetric and the center position is consistent with the optical 
center within
the {\it Chandra}'s pointing accuracy. This feature also suggests that
evaporation or stripping of the ISM is not an ongoing process now.

\citet{vik01} discuss  the thermal balance between radiative
cooling and thermal conduction from the surrounding ICM, based on the
{\it Chandra} data for the Coma galaxies. The ISM temperature of NGC 4889
($kT\sim 1.8 $ keV) was too high compared with the gravitational potential
of the galaxy, and a positive temperature gradient was observed in NGC
4874. These features suggested the importance of heat input from the
surrounding ICM\@. In the A1060 galaxies, the conductive heat flux
from the ICM would be smaller because of much lower ICM temperature
($kT=3.1$ keV compared with $\sim 8$ keV in the Coma cluster).  The
temperature gradient is not clearly recognized in the ISM, although
the poor statistics allow some temperature increase with radius.
The ISM temperature can
also be evaluated from the $\sigma^{2}-kT$ relation \citep{mat01}.  
The velocity
dispersions are 187 km s$^{-1}$ for NGC 3311 and 236 km s$^{-1}$
for NGC 3309 \citep{pru96}, respectively, leading to 
$\beta_{\rm spec}=\mu m_{p} \sigma^{2}/kT$ values of 0.252 and 0.453.
Since typical $\beta_{\rm spec}$ values within $1.5 r_{e}$ are 0.5--1
\citep{mat01,hiro97}, the ISM in NGC 3311 and NGC 3309 is hotter than
the kinetic temperature of the stars.  Since heat conduction at the
outer edge of the ISM is likely to be suppressed from the evaporation
argument, the high temperatures in the A1060 galaxies would be
the result of  adiabatic compression of the ISM in the cluster.

Now, let us consider about the dynamical state of the ISM in NGC 3311 
and NGC 3309. 
As shown above, even though the radiative cooling times of several 
times 10$^{8}$ years 
are short, inputs from stellar mass loss and Type-1a supernova can
maintain the present amount of the ISM\@.
The low $\beta_{\rm spec}$ values  suggest that 
the hot ISM have sufficient energy to escape from the 
galaxy potential, even though there is no morphological sign of 
galactic winds. A slow evaporation of the ISM due to the suppressed 
conductive heat input from the ICM may be the actual process going
on.
Regarding the large scatter of the $L_{X}/L_{B}$ ratios, models
incorporating partial winds have been recently developed (e.g.\ 
Pellegrini \& Ciotti 1998).
According to the model, the change of stagnation radius 
between the outflowing outer region and 
the inflowing inner region controls the X-ray luminosity.
However, these models consider isolated systems and the effect of 
external pressure needs to be further studied.

The {\it Chandra} observation of the central region of A1060 has
confirmed that the cluster is indeed the archetype
isothermal cluster under hydrostatic equilibrium. When the
ISM emission is subtracted, the ICM emission is shown to be very isothermal
with no sign of cool component, even though the systematic
uncertainty in the soft band response hampers us to set a meaningful
upper limit. The X-ray size of the ``cD'' galaxy NGC 3311 is very
compact with no extended potential structure, which has been commonly 
seen in X-ray
bright elliptical galaxies \citep{mat01}. Morphological study of a large
sample of PSPC images, on the other hand, showed that the cooling 
flow clusters
tend to indicate more circularly symmetric shapes than the systems
accompanied by radio halos and relics \citep{sch01}. In this respect,
A1060 seems to be standing at a subtle location in the evolution
process. Numerical simulations suggest that cooling flows can be
disrupted by major mergers \citep{mcg84} and a highly inhomogeneous
temperature distribution should remain for about 2 Gyr \citep{roe97}.
A possible interpretation for A1060 is that a sufficiently long time
($\gg 2$ Gyr) has passed from the last major merger, which has stirred
the central ICM, so that good isothermality and hydrostatic
equilibrium was reached. However, the elapsed time is not long enough
to assemble the matter and to establish a hierarchical potential
structure dominated by the cD galaxy. This picture is also consistent
with the observed uniform distribution of metals in the A1060 ICM
\citep{tam00,oha01}.

\section{Conclusion}

  {\it Chandra} observation of the central region of the relaxed cluster
of galaxies, A1060, confirmed that the temperature of the ICM was
almost constant from outside to the inner 20 kpc region. The X-ray 
emission from
the ISM of the two giant elliptical galaxies, NGC 3311 and NGC 3309,
were clearly resolved. X-ray properties of the two galaxies are very
similar and the ISM morphologies do not show a stripping feature even though
the
velocity difference between the two galaxies is $\sim 500$ km s$^{-1}$.  
The ISM
temperatures are constant at 0.7--0.9 keV, and the X-ray to
optical luminosity ratios for the sum of the ISM and LMXB components are
within the scatter for other galaxies.  The spatial
extents of the ISM emission are, however, as small as $2 - 3$ kpc for
both galaxies, and the pressure balance between the ISM
and the ICM is achieved at the ISM boundaries.  We discuss, based on these
results, that the
ISM is mainly supplied from stellar mass loss and is confined by
the external pressure of ICM, with highly suppressed heat
conduction. The ICM in the central region of A1060 is relaxed,
and the extended potential structure around the cD galaxies has not grown
up yet, possibly due to gas mixing in past merger episodes.

\acknowledgments
The authors thank the referee, and Dr.\ K. Matsushita for providing valuable comments 
and for support in using the data in Figure 6.
T. F. is supported by the Japan Society for the 
Promotion of Science Postdoctoral Fellowships for Research Abroad.
This work was partly supported by the Grants-in Aid for Scientific 
Research No. 12304009 and No. 12440067 
from the Japan Society for the Promotion of Science.
This research has made use of the NASA/IPAC 
Extragalactic Database (NED) which is operated
by the Jet Propulsion Laboratory, California Institute 
of Technology, under contract with the 
National Aeronautics and Space Administration.
The Digitized Sky Survey was produced at the Space Telescope Science
Institute under U.S. Government grant NAG W-2166.

\clearpage


\clearpage
\begin{deluxetable}{lllllll}
\tablecaption{Resultant parameters of $\beta$ model fits for radial profiles 
\label{tab:beta}}
\tablewidth{0pt}
\tablehead{
\colhead{object} &
\colhead{fit radius\tablenotemark{a}}&
\colhead{$\beta$} &
\colhead{core radius\tablenotemark{a}} &
\colhead{normalization\tablenotemark{b}} &
\colhead{background\tablenotemark{b}} &
\colhead{$\chi^{2}$/dof} 
}
\startdata
ICM\tablenotemark{c} & 294& 0.36$\pm0.02$& 100.9$_{-10.6}^{+12.5}$ &10.2$\pm$0.37  & \nodata & 69.8/58 \\
NGC 3311\tablenotemark{c}& \nodata & 0.71$_{-0.16}^{+0.41}$& 3.95$_{-1.44}^{+2.63}$ &97.5$_{-25.2}^{+33.9}$ & \nodata& \nodata\\
NGC 3311  & 49& 0.67$_{-0.14}^{+0.30}$   & 3.79$_{-1.37}^{+2.26}$ &95.4$_{-24.4}^{+15.7}$ & 9.34$_{-0.45}^{+0.33}$& 96.9/96\\
NGC 3309  & 49& 1.08$_{-0.27}^{+0.85}$   & 3.73$_{-1.10}^{+2.30}$ &220.6$\pm4.95$ & 6.81$\pm$0.25 & 95.6/96 \\
\enddata
\tablenotetext{a}{arcsec}
\tablenotetext{b}{10$^{-4}$ c arcsec$^{-2}$ cm$^{-2}$}
\tablenotetext{c}{Based on a double-$\beta$ fit assuming the ICM and 
NGC 3311 emission together}
\tablecomments{The energy range is between 0.5 and 1.5 keV\@. 
Errors denote the 90\% confidence limits for a single parameter.}
\end{deluxetable}

\begin{deluxetable}{lllll} 
\tablecaption{Obtained temperature  of  NGC 3311 and  NGC 3309.
\label{tab:spec}}
\tablewidth{0pt}
\tablehead{
\colhead{object} &
\colhead{$kT$ (keV)} &
\colhead{$F_{X}$(ISM)\tablenotemark{a}} & 
\colhead{$F_{X}$(LMXB)\tablenotemark{a}} &
\colhead{$\chi^{2}$/dof}
}
\startdata
NGC3311& 0.87$_{-0.09}^{+0.21}$ &3.9(2.9-5.6) &4.5(3.1-7.2) & 23.2/25 \\
NGC3309& 0.77$_{-0.07}^{+0.10}$ &2.6(2.2-3.7) &3.7(1.9-4.2) & 44.7/35 \\
\enddata 
\tablenotetext{a}{Flux unit: 10$^{-14}$ erg cm$^{-2}$s$^{-1}$ between 0.5 and 4.5 keV }
\tablecomments{
Errors denote the 90\% confidence limits for a single-parameter. 
The fitting model is wabs(mekal+zbrems), in which fixed parameters are 
the column density $N_{\rm H}$ 
at the Galactic value of 6\pten{20}cm$^{-2}$, the redshift $=0.0114$, 
the abundance at 0.5 solar, and the temperature of zbrems 
at $kT =10$ keV\@.}
\end{deluxetable}

\end{document}